\documentclass[aps,epsfig,float,twocolumn]{revtex4}
\newcommand{\be}{\begin{equation}}
\newcommand{\ee}{\end{equation}}

\usepackage{graphicx}
\usepackage{amsmath}
\usepackage{amstext}
\usepackage{amssymb}
\usepackage{amsfonts}
\usepackage{amsbsy}

\begin{document}
\title{Solving Gapped Hamiltonians Locally}
\author{M. B. Hastings}
\affiliation{Center for Nonlinear Studies and Theoretical Division,
Los Alamos National
Laboratory, Los Alamos, NM 87545
}
\email{hastings@lanl.gov} 

\date{August 23, 2005}
\begin{abstract}
We show that any short-range Hamiltonian with a gap between the ground
and excited states
can be written as a sum of local operators, such that
the ground state is an approximate eigenvector of each operator
separately.
We then show that the
ground state of any such Hamiltonian is close to a generalized
matrix product state.  The range of the given operators needed to
obtain a good approximation to the ground state
is proportional to the square of the logarithm of the system size times a
characteristic
``factorization length".
Applications to many-body quantum
simulation are discussed.
We also consider density
matrices of systems at non-zero temperature.
\vskip2mm
\end{abstract}
\maketitle

The application of numerical renormalization group to quantum systems is a
natural idea with a long history.  Despite Wilson's success with the
Kondo model\cite{wilson}, other early attempts based on keeping low-lying
eigenstates in each block were less successful\cite{early}.  The
basic idea of these methods is to break a system into subsystems,
solve each of the subsystems, and then join the solutions
together.

This leads to the following general question: how do you solve a
system if you know the solution of its subsystems?  Consider, for example,
the following toy impurity problem: a single spin-$1/2$ impurity embedded
in a band gap insulator.  Suppose that the electrons in the insulator do
not interact with each other, but only with the impurity spin.  The
problem of the electrons alone can readily be solved, even in an
infinite system, by filling Bloch
states up to the Fermi level, and one finds exponentially decaying
correlations in this system.  If the interaction with the impurity is
strong, the impurity problem does not
admit an analytic solution, but given the finite correlation length,
one might approximately
solve a finite periodic region around the impurity on a computer.
How, though, can one write down a solution for the combined system?
How can one ``sew" the two solutions together?

In one-dimension, density matrix renormalization group (DMRG)\cite{dmrg}
provides a means to do exactly this and has been
extremely successful.  The states that it finds are
matrix product states\cite{mprod,mp1,mp2}.  There exist some
promising higher dimensional generalization of
these matrix product states\cite{mprod,2d}.
Precise bounds on how well one can approximate a given
quantum state by a matrix product state are still lacking, however.

Recently, there have been several advances in understanding the connection
between a gap and the locality of correlation
functions\cite{lsmh,loc1,loc2,topo1,wen},
providing a firm analytical basis for the notion that a gap
implies exponentially decaying correlations while a power law density
of states implies correlations are bounded by an algebraic decay.
In this paper, we will
similarly use the existence of a gap to study this problem of sewing states
together.

All of the known Hamiltonians that give matrix product states, such
as the AKLT\cite{aklt} Hamiltonian, have the property that the ground state
is an eigenvector of each term in the Hamiltonian separately.
We refer to this as a local projective Hamiltonian.
The first portion of this paper will 
construct a form of arbitrary gapped Hamiltonians such that
the ground state is an {\it approximate} eigenvector of each term separately,
writing the Hamiltonian as a sum of terms in Eq.~(\ref{hsum}).
This will be  a first step to building a matrix product form of the
ground state.  We construct such a form in this paper, but do not bound
the number of states required in the matrix product construction.  Such
a bound will be given in a future work.
While the proofs here will be to some extent constructive, they will assume that
certain properties of the ground state are known, and thus they are not
so useful in themselves for the problem of finding ground states.

Next, we briefly discuss applications to numerical
simulation.  Of course, one application is in analyzing existing algorithms,
but we consider suggest the possibility of
different algorithms based on the proofs in the first
part.  In this case, we will discuss how to find the needed properties
of the ground state used in the proofs in the first part.

The last portion of the paper will consider systems at non-zero temperature.
In this case we will show that the density matrix of the system can
be written in a matrix product form, which provides a higher dimensional
generalization of the one-dimensional matrix product form for density
matrices\cite{1dft}.

\section{Approximate Local Projective Form of the Hamiltonian}
Consider the AKLT Hamiltonian\cite{aklt}, $H=\sum H_i$, with
$H_i=\vec S_i \cdot \vec S_{i+1}+
(1/3)(\vec S_i \cdot \vec S_{i+1})^2$.  This Hamiltonian has an exact
matrix product ground state.  For a chain of $N-1$ sites, suppose there
are ground states labeled by an index $\beta$.  Then, a chain of
$N$ sites is supposed to have ground states labeled by an index $\alpha$,
with $\alpha\rangle=\sum_{\beta,s} A_{\alpha,\beta}(s) 
\beta\rangle \otimes s\rangle$,
where $s\rangle$ denotes a complete set of states on
the $N$-th site (in this case, there are three such states), and
$A_{\alpha,\beta}(s)$ is the matrix defining the matrix product state.
This then gives a wavefunction which sews the solutions of the two subsystems
together.  One way to find such wavefunctions is DMRG, while another
is the variational matrix product method.  For the AKLT chain,
the ground state has $\alpha=1,2$ with 
\be
\sum_s A_{\alpha,\beta}(s) s\rangle=
\begin{pmatrix}
0\rangle & i\sqrt{2} \, +\rangle \\ i\sqrt{2} \, -\rangle & 
- \, 0\rangle
\end{pmatrix}.
\ee
For more general Hamiltonians, the range of the indices $\alpha,\beta,s$
may be larger and the matrix $A_{\alpha,\beta}(s)$ may be different.

This ground state not only minimizes the Hamiltonian $H$, but also
minimizes each $H_i$ individually, and thus is an eigenvector
of each $H_i$.  This observation is fundamental
to the work in this section.  We take an arbitrary
Hamiltonian and approximately
rewrite it in an {\it approximate local projective form},
defined to be
a form in which the Hamiltonian is a sum of local terms $M_i$ such that
the ground state is close (as defined below) to an eigenvector of
each $M_i$ separately.
A Hamiltonian with such a form can truly be solved by solving
subsystems separately.  Break a chain of $N$ sites up into two
subchains of $N-m$ and $m$ sites.  For each chain, find the eigenvectors of
the Hamiltonian with the correct eigenvalue (the same eigenvalue as
the ground state of the full system has for the Hamiltonian of the subchain).
The ground state of the full chain will be a linear
combination of outer products of the given states in each subchain.  The
matrix product method realizes this for $m=1$.

We consider an arbitrary Hamiltonian $H$ which is assumed to have some
number of degenerate
ground states and then a gap $\Delta E$ to the rest of the
spectrum.
That is, define $\Psi_a\rangle$ to be an eigenstate
state of $H$ with energy $E_a$, and let $\Psi_a\rangle$ for $a=0...n-1$ be $n$
distinct
ground states while for $a\geq n$ we have $E_a\geq \Delta E$.
We consider the case $E_0=E_1=...E_{n-1}=0$\cite{split}.
We assume that $H$ obeys the finite range conditions\cite{fgv,lsmh}.
That is, the Hamiltonian $H$ can be written as a sum of terms $H_i$ such
that each $H_i$ has a bounded operator norm, $||H_i|| \leq J$ for some $J$
and such that each $H_i$ acts only on sites within some interaction 
range $R$ of site $i$\cite{exp}, while there are at most $S$ sites $j$
within distance $R$ of sites $i$ for any $i$.
Introduce some metric on the lattice $d(i,j)$
as the distance between sites $i$ and $j$, while $d(O,j)$ is defined
to be the distance between an operator $O$ and a site $j$: that is,
the minimum, over sites $i$ on which operator $O$ acts, of 
$d(i,j)$.
Then,for any operator $O_j$ which acts on a site $j$ with $d(i,j)>R$,
we have $[H_i,O_j]=0$.

We now construct the approximate local projective form, and in
the next section construct the ground states.  Following\cite{loc1},
define 
\be
\label{tnd}
\tilde H_i^0=\frac{\Delta E}{\sqrt{2 \pi q}} \int_{-\infty}^{\infty}
{\rm d}t
\tilde H_i(t),
\ee
where
\begin{eqnarray}
\label{thd}
\tilde H_i(t)\equiv H_i(t) \exp[-(t\Delta E)^2/(2 q)],\\ \nonumber
H_i(t)=\exp(iHt) H_i \exp(-iHt),
\end{eqnarray}
with $q$ to be chosen later.
That is, $H_i(t)$ is defined following the usual Heisenberg evolution
of operators, while $\tilde H_i(t)$ is equal to $H_i(t)$ multiplied
by a Gaussian which cuts off the integral in Eq.(~\ref{tnd}) at times $t$
of order $\sqrt{q}/\Delta E$.  The notation of \cite{loc1} for
these operators $\tilde H_i^0$ is chosen
to indicate that we make an approximation (hence the tilde) to the
zero frequency (hence the zero) part of the $H_i$.

Now, here is the key point of the paper.  We claim that $\tilde H_i^0$,
acting on a ground state, gives back another ground state up to some
exponentially small difference.
To see this, compute
$(\tilde H_i^0)_{ab}\equiv\langle \Psi_a,\tilde H_i^0 \Psi_b \rangle$,
the matrix element of $\tilde H_i^0$ between
states $\langle\Psi_a$ and $\Psi_b\rangle$\cite{notation}.
A direct computation gives
\be
\label{mel}
(\tilde H_i^0)_{ab}=(H_i)_{ab} \exp[-q(\frac{E_a-E_b}{\Delta E})^2/2].
\ee
Let \be
P_{low}=\sum_{0\leq a \leq n}\Psi_a\rangle\langle\Psi_a
\ee
and
\be
P_{high}=1-P_{low}.
\ee
Thus, $P_{\rm low}$ projects onto the space of
ground states while $P_{\rm high}$ projects onto the remaining states.
Then, from Eq.~(\ref{mel}), the norm 
\be
\label{gs}
|P_{high} \tilde H_i^0\Psi_a\rangle|\leq
||H_i^0|| \exp(-q/2)\leq J\exp(-q/2),
\ee
as claimed.

One important fact is that
\be
\label{hsum}
\sum_i \tilde H_i^0=\sum_i H_i=H.
\ee
Also, $H_i$ and $\tilde H_i^0$
have the same matrix elements in the subspace of ground states:
$(\tilde H_i^0)_{ab}=(H_i)_{ab}$ if $0\leq a \leq n$ and $0\leq b \leq n$.
Finally,
\be
\nonumber
||H_i^0|| \leq ||H_i|| \leq J. 
\ee 

The $\tilde H_i^0$ are {\it local} in that the commutator
of $H_i$ with any operator $O_j$ which acts only on a site
$j$ is exponentially small in $d(i,j)$.  This follows
since Eq.~(\ref{tnd}) defines $\tilde H_i^0$ as an integral of $H_i(t)$ over
times $t$; the Gaussian in Eq.~(\ref{thd}) cuts this
integral off for sufficiently long time while for short time
$H_i(t)$ is local.  
The precise statement shown in the Appendix is that, for any
operator $O_j$ which acts only on a site $j$,
\be
\label{lbnd}
||[\tilde H_i^0,O_j]|| \leq
J ||O_j|| (g(c_1 l,l) +
2 \exp[-(c_1 l \Delta E)^2/(2 q)]),
\ee
where $l=d(H_i,j)$ and
the function $g(c_1 l,l)$ is an exponentially decaying function
of $l/\xi_C$ for some microscopic length scale $\xi_C$ of order
the interaction range $R$.  The
constant $c_1$ is a characteristic inverse velocity of propagation
in the system; the existence of a finite velocity of propagation, as
discussed in the Appendix, is essential in showing that
$H_i(t)$ is local for short time.  Eq.~(\ref{lbnd}) implies that
$\tilde H_i^0$ is local in that it has a small commutator
with operators which are far enough from $i$.

We can further define $M_i$ to be
an approximation to $\tilde H_i^0$ which is truly finite range:
$M_i$ will exactly commute with $O_j$ if $d(i,j)$ is greater
than a certain range $l_{proj}$.
To do this, define 
\be
\label{mid}
M_i=(\Delta E/\sqrt{2 \pi q}) \int {\rm d}t
\exp[-(t\Delta E)^2/2] H_i^{\rm trunc}(t),
\ee
where 
\begin{eqnarray}
\label{odef}
H_i^{\rm trunc}(t)=\exp(i H_{\rm loc} t) H_i \exp(-i H_{\rm loc} t), \\
\label{hpdef}
H_{\rm loc}=\sum_{j,d(i,j)\leq l_{proj}-R} H_j.
\end{eqnarray}
Thus, $H_{\rm loc}$ is the sum
of terms $H_j$ with $d(i,j)$ less than $l_{proj}-R$, so that
$H_i^{\rm trunc}(t)$ only acts on sites within distance $l_{proj}$ of $i$.
Thus, the procedure to define the $M_i$ is very simple: one
uses the definition Eq.~(\ref{thd}), but as one evolves $H_i(t)$
one drops terms which involve sites more than $l_{proj}$ from $i$.

In the Appendix, we show that
\begin{eqnarray}
\label{apb}
||\tilde H_i^0-M_i|| &\leq&
J (N(l_{proj}) g(c_1 l_{proj},l_{proj}) 
\\ \nonumber
&+&
2 \exp[-(c_1 l_{proj} \Delta E)^2/(2 q)]),
\end{eqnarray}
where $N(l_{proj})$ is defined to be
the number of sites $j$ with
$l_{proj}-R<d(i,j)\leq l_{proj}+R$.  Note also that $||M_i||\leq J$.

We now pick $q$.  For a given range $l_{proj}$ of the $M_i$, we
want to minimize
$|P_{high} M_i\Psi_a\rangle|$, so that the ground states
are approximate eigenstates of the $M_i$.  
By a triangle inequality,
\begin{eqnarray}
\nonumber
|P_{high} M_i\Psi_a\rangle|
&\leq&
|P_{high} \tilde H_i^0 \Psi_a\rangle|+
||\tilde H_i^0-M_i||\\ \nonumber
&\leq&
J(\exp(-q/2)+
N(l_{proj}) g(c_1 l_{proj},l_{proj})\\ \nonumber
&+&
2\exp[-(c_1 l_{proj} \Delta E)^2/(2 q)]).
\end{eqnarray}
To get the best bound,
we pick $q=c_1 l_{proj}\Delta E$.  Then,
\be
\label{facld}
|P_{high} M_i \Psi_a \rangle|\leq
J {\cal O}(\exp(-l_{proj}/l_{fac})),
\ee
where ${\cal O}$ denotes a quantity of order $\exp(-l_{proj}/l_{fac})$,
with $l_{fac}$ being the characteristic {\it factorization length}.
The length $l_{fac}$ is equal to the minimum of $(c_1 \Delta E)^{-1}$ and
$\xi_C$, and thus for small $\Delta E$, $l_{fac}=(c_1 \Delta E)^{-1}$.

With the given $q$, the bound in Eq.~(\ref{apb}) becomes
\begin{eqnarray}
\nonumber
||\tilde H_i^0-M_i|| &\leq&
J [N(l_{proj}) g(c_1 l_{proj},l_{proj}) \\ \nonumber
&+&
2 \exp(-c_1 l_{proj} \Delta E/2)].
\end{eqnarray}
This difference is exponentially
small in $l_{proj}/l_{fac}$, so that difference between the ground state
energy per site of $H=\sum_i H_i$ and that of the Hamiltonian
$M=\sum_i M_i$ is exponentially small in $l_{proj}/l_{fac}$.  Defining
$N$ to be the number of sites $i$ in the system, if $N ||\tilde H_i^0-M_i||$
is less than of order $\Delta E$, then the ground state of $M$ has a
non-vanishing
projection onto the ground state of $H$.  This requires an $l_{proj}$ which
is of order $\log(N)$.

We claim that these $M_i$
realize the approximate local projective
form, 
\be
H \approx M=\sum_i M_i.
\ee
We start with the simplest case of only one ground state, $n=1$.
Then, Eq.~(\ref{facld}) implies that the ground state
$\Psi_0\rangle$ is close to an eigenvector
of each $\tilde M_i$.  That is,
\be
\nonumber
|M_i \Psi_0 \rangle - \langle M_i \rangle \Psi_0 \rangle|\leq
J {\cal O}(\exp(-l_{proj}/l_{fac})),
\ee
where $\langle ... \rangle$ denotes the
ground state expectation
value.  By picking $l_{proj}$ large, we can make this difference as small as
desired.

The $M_i$ give the desired approximate local projective form for the case
of a unique ground state.  We will show in the next section how to
construct matrix product states that are approximate ground states
of this Hamiltonian.  However, we first consider
the case of multiple degenerate ground states.

If there are multiple ground states with a gap to the rest of the
spectrum, then the situation is slightly more complicated.
Eq.~(\ref{gs}) implies that the $\tilde H_i^0$ acting
on ground states gives states which are close to ground states, but no
longer necessarily
implies that the ground states are eigenstates of the $\tilde H_i^0$.
Instead, it depends to some extent on what basis we choose for the ground
states.  As a simple example, consider the Majumdar-Ghosh Hamiltonian for
a one-dimensional spin-$1/2$ chain: $H=\sum_i H_i$ with
$H_i=J \sum_i [\vec S_i \cdot \vec S_{i+1}+
(1/2) \vec S_i \cdot \vec S_{i+2}]$.  This Hamiltonian has two exact
ground states; in one state
sites $i=1,2$ are in a singlet, sites
$i=3,4$ are in a singlet, and so on, while in the other state 
sites $i=2,3$ are in a singlet, sites $i=4,5$ are in a singlet, and so on.
Denote the first state by $\Psi_{\rm even}\rangle$ and the second state
by $\Psi_{\rm odd}\rangle$.  
Now, the states $\Psi_{\rm even}\rangle$
and $\Psi_{\rm odd}\rangle$ break translational symmetry; the expectation
value
$\langle \Psi_{\rm even}, H_i \Psi_{\rm odd} \rangle$ is an alternating
function of $i$.  However, in an infinite system the expectation value
$\langle \Psi_{\rm odd}, H_i \Psi_{\rm odd} \rangle$ vanishes.
Thus, in the subspace formed by the two vectors
$\Psi_{\rm even,odd}\rangle$, the $H_i$ are diagonal and therefore the
$\tilde H_i^0$ are also diagonal in this subspace,
So with the given $M_i$, there is no problem in this basis
of ground states: the states $\Psi_{\rm even,odd}\rangle$ are approximate
eigenstates of the $M_i$ for large $q$.
Of course, as is well known for
the Majumdar-Ghosh chain, if we were to pick $H_i=(J/2) 
[\vec S_i \cdot \vec S_{i+1}+
\vec S_i \cdot \vec S_{i-1}+
(1/2) \vec S_{i-1} \cdot \vec S_{i+1}]$, then the states
$\Psi_{\rm even,odd}\rangle$
would be exact eigenstates of the $H_i$, but let us suppose that we do not
know that this form of the Hamiltonian is available.

Suppose instead we choose to form ground states which are
eigenvectors of the translation operator by
$\Psi_S\rangle=\Psi_{\rm even}\rangle+\Psi_{\rm odd}\rangle$ and
$\Psi_A\rangle=\Psi_{\rm even}\rangle-\Psi_{\rm odd}\rangle$.
Then, the $H_i$ are {\it not}
diagonal in this subspace, and the states $\Psi_{S,A}\rangle$ are not
approximate eigenvectors of the $\tilde H_i^0$, no matter how large $q$ is.
One way to get around this is to 
go to an enlarged unit cell of two sites, setting
$H=\sum_j H_j'$ where $H_j'=H_{2j}+H_{2j+1}$ and then
the states $\Psi_{S,A}\rangle$ are approximate eigenvectors of $M_j'$.
However, the simplest solution is to use the states $\Psi_{\rm even,odd}\rangle$
instead of $\Psi_{S,A}\rangle$.

Thus, the important question is: 
can we simultaneously diagonalize
all of the $M_i$ in the subspace formed by the ground states $\Psi_a\rangle$ for
$0 \leq a \leq n$?  If so, then we can ensure that, by the appropriate
choice of basis for the ground states, each ground state is an
approximate eigenvector of each of the $M_i$ and we will have
\be
\label{low}
| (M_i-\langle \Psi_a , M_i \Psi_a \rangle) \Psi_a \rangle|\leq
J {\cal O}(\exp(-l_{proj}/l_{fac})),
\ee
If the states $\Psi_a\rangle$
in this basis break translational symmetry, then the expectation
value of $M_i$ in a state $\Psi_a\rangle$ may depend on $i$ even
if $H$ is translationally invariant.  
In order to simultaneously
diagonalize the $M_i$ in the subspace of ground states,
we need the $M_i$ to commute in this subspace.

The conditions for the $M_i$ to commute in this subspace
are discussed in an Appendix.  We will show that,
except for a few artificial examples, the $M_i$ approximately
commute.  In particular,
if $H$ is translationally invariant, we will show that the commutator
of the $M_i$ in this subspace vanishes exponentially
in the system size, and thus we can pick $\Psi_a\rangle$ such that
Eq.~(\ref{low}) holds.
For $H$ which are not translationally
invariant, we show that most of the $M_i$ commute in this subspace.
More precisely, for any basis of the ground states, define $o(i)$ to be the
operator norm of the off-diagonal part of $M_i$ in the low energy
sector in that basis (the off-diagonal part of $M_i$ is a matrix
which has zeros on the diagonal, but whose off-diagonal elements are
the same as $M_i$).
We will show in the Appendix how choose a basis in which
we can bound $\sum_i o(i)$ by Eq.~(\ref{oibu}).
This implies
that we can pick the $\Psi_a\rangle$ such that 
\begin{eqnarray}
\nonumber
|M_i \Psi_a \rangle - \langle \Psi_a , M_i \Psi_a \rangle \Psi_a \rangle|\leq
J {\cal O}(\exp(-l_{proj}/l_{fac}))+o(i),
\end{eqnarray}
with the sum of
$o(i)$ bounded.

\section{Matrix Product State for the Local Projection Hamiltonian}
We pick a given ground state $\Psi_a\rangle$ and approximate it
by a matrix product state, with an appropriately bounded error.
For simplicity, we consider the case in which we can pick
the $\Psi_a\rangle$ such that Eq.~(\ref{low}) holds.
This includes, as discussed
above, all translationally invariant systems, as well as many
translationally invariant systems, those without local zero energy
excitations as discussed in the Appendix.  
We review previous work on matrix product or
valence bond states, and then provide various constructions of the matrix
product state in the given case. 
The value of $l_{proj}$ required to obtain a good approximation
to the ground state will be seen to grow as a power of the logarithm of the
system size.

\subsection{Matrix Product and Valence Bond States}
The one-dimensional matrix product state, as discussed above, takes
a chain of $N-1$ sites with a set of ground states $\beta$, and
constructs the ground states of an $N$ site chain by 
$\alpha\rangle=\sum_{\beta,s} A_{\alpha,\beta}(s) s\rangle \times
\beta\rangle$, where $\beta\rangle$ are ground states of an $N-1$ site chain.  
Matrix product ground states in one-dimension can always be written as
valence bond states\cite{mprod}.  Construct
an enlarged Hilbert space on each site, labeling states on site
$i$ by two indices $\alpha_i,\beta_i$.  A wavefunction is
constructed in the enlarged space, such that this wavefunction
$\psi(\alpha_1,\beta_1,\alpha_2,\beta_2,...)$ is a product of wavefunctions
$\psi(\beta_1,\alpha_2) \psi(\beta_2,\alpha_3),...$.  Finally,
a map is written on each site from the original Hilbert space $s\rangle$
to the enlarged Hilbert space $\alpha_i\beta_i\rangle$, and the
wavefunction on the original Hilbert space is defined to be the
wavefunction of the mapped state on the enlarged Hilbert space.
To make this concrete, suppose the matrix product construction gives
$\alpha\rangle=\sum_{\beta,s} A_{\alpha,\beta}(s) s\rangle \times
\beta\rangle$.  Then, set
$\psi(\beta_i,\alpha_{i+1})=\delta(\beta_i,\alpha_{i+1})$.
Let $F$ map state $s_i\rangle$ on site $i$ onto
$\sum_{\alpha_i\beta_i} A_{\alpha_i,\beta_i}(s_i) \alpha_i\beta_i\rangle$.
Alternately, this map $F$ can be viewed as a projection
from the space of states $\alpha_i\beta_i\rangle$ onto the states
$s_i\rangle$\cite{peps}.
Then, the matrix product wavefunction is given by
$\psi(F(s_1,s_2,...))$.

In the AKLT case,
$\alpha_i$ labels one of the two spin-$1/2$s and $\beta_i$ labels the other
one.  The product wavefunction is a product of singlet pairs, while
the map $F$ projects the two spin-$1/2$s onto a spin-$1$.
The matrix product state for the AKLT chain can be written
also as a valence bond state.
Each site has a spin-$1$, which may be represented by two spin-$1/2$
spins.  One spin-$1/2$ is in a singlet with a spin-$1/2$ on the next site
to the right, and one is in a singlet with a spin-$1/2$ on the next site
to the left.  This state is then projected onto the spin-$1$ state of the
two spin-$1/2$ spins on each site.
                                                                                
Matrix product states and valence bond states
are equivalent.  However, the discussion above was
confined to pure states (wavefunctions) on finite systems.
In \cite{mprod}, matrix product and valence bond states were
also constructed for mixed states (density matrices) and again
shown to be equivalent.

Thus far we have discussed systems in one dimension.  A higher dimensional
system can always be viewed as a one dimensional system as follows.  For
a $d$-dimensional system in which each site is labeled by $d$ coordinates,
all of the sites with a particular value of one coordinate can be grouped
into one supersite, leaving a one dimensional chain.  This method is very
limited
in practice; for a system of linear size $L$, the size of the Hilbert space
on a single supersite is exponential in $L^{d-1}$ and thus the
range of the indices $\alpha,\beta$ is also exponentially large.
This method amounts to
studying a one-dimensional ladder system.
                                                                                
Valence bond states are often regarded as a more
appropriate way for constructing states in more than one dimension.
To construct
such a state\cite{mprod,2d}, on an arbitrary lattice in any number
of dimensions, for each site $i$ one constructs one $k$-dimensional
auxiliary Hilbert space per bond, where the bond connects
site $i$ to site $j$.  An wavefunction is defined in the enlarged
Hilbert space, which is a product of wavefunctions on each bond, where
the wavefunction on each bond is a function of states at the ``ends" of the
bond (in one dimension, these are the indices $\beta_i,\alpha_{i+1}$).
For each site, a map is defined from the
original Hilbert space to the product of the auxiliary Hilbert spaces
on that site.  In this paper, we restrict to methods based on supersites
for higher dimensions, while a future publication will provide valence
bond constructions in this case\cite{tbp}.
                                                                                
Manipulating these higher dimensional valence bond states is difficult.
After \cite{mprod}, most higher dimensional work involved special examples
where the Hamiltonian was exactly equal to a sum of projection
operators\cite{highd}, so that the ground state is exactly
a higher dimensional generalization of the AKLT ground state.
However, in an important advance\cite{2d},
valence bond states were suggested as a good ansatz
for arbitrary Hamiltonians, with
a numerical technique being used to compute the state.  We now provide
a construction of the matrix product or valence bond states.

\subsection{Construction of Matrix Product State}
We first give the matrix product construction in one dimension.  The
approach discussed above in one dimension is an iterative procedure: from
the ground states of an $N-1$ site chain,  we construct those of an
$N$ site chain.  This procedure is shown schematically in Fig.~1(a).  We
first find the set of allowed states on the first two sites, then the
set of allowed states on the first three sites, and so on.  Our procedure
will instead be a ``hierarchical" procedure, as shown in Fig.~1(b).  We will
find a set of allowed states on pairs of sites; we then find sets of allowed
states on groups of four sites, and so on.  In practice in DMRG, the iterative
approach works better, but we find the hierarchical approach gives better
bounds here.

Consider
a one-dimensional system with given $l_{proj}$.  
First group the sites into supersites with
sites $1\leq i \leq 2 l_{proj}$ grouped into one supersite, sites
$2 l_{proj}+1\leq i \leq 4 l_{proj}+1$ into another supersite,
and so on, grouping $2 l_{proj}$ sites
into each supersite.  
Suppose each site has an $m$-dimensional Hilbert space.
The dimension of the space of states on the sites 
$1\leq i \leq 2 l_{proj}$ is at most $m^{2l_{proj}}$.  
From now on, we refer to these supersites simply as ``sites";
with this grouping, the operators $M_i$
act only on pairs of neighboring sites, with no longer range interaction or
three site interactions.
We let
$M_{k,k+1}$ be the sum of the operators $M_i$ which act on sites $k$ and $k+1$.
Label the
supersites of the system by $i=1...N$.  The system is periodic, so that
site $N+1$ is identical to site $i$; similarly,
$d(N,1)=1, d(N,2)=2$, and so on.

Pick a given $\Psi_a$.
Let $\alpha^1_k$ label a complete basis of states $\alpha^1_k\rangle$
on site $k$.
From Eq.~(\ref{low}),
\begin{eqnarray}
\nonumber
|(M_{k,k+1}-\langle \Psi_a, M_{k,k+1} \Psi_a \rangle)
\Psi_a\rangle| \\ \nonumber
\leq J l_{proj} {\cal O}(\exp(-l_{proj}/l_{fac})).
\end{eqnarray}
Let $P^1_k$ project onto the eigenvectors of $M_{k,k+1}$ with eigenvalues
$\lambda$ such that $|\lambda-\langle \Psi_a,M_{k,k+1} \Psi_a \rangle|\leq
x$, for some $x$ to be chosen later.
Let $\alpha^2_k$ label these different eigenvectors $\alpha^2_k\rangle$.
Then, 
\begin{eqnarray}
\nonumber
x \langle \Psi_a,(1-P_k^1) \Psi_a \rangle
&\leq& |(M_{k,k+1}-\langle \Psi_a, M_{k,k+1} \Psi_a \rangle)
\Psi_a\rangle| \\ \nonumber &\leq& J l_{proj} {\cal O}(\exp(-l_{proj}/l_{fac})).
\end{eqnarray}
Then,
choose $x=J l_{proj} {\cal O}(\exp[-l_{proj}/(\log_2(N) l_{fac})])$, so that
\be
\langle \Psi_a, P^1_k \Psi_a \rangle \geq 1-b_1,
\ee
where
$b_1=\exp[-(1-1/\log_2(N))(l_{proj}/l_{fac})]$.
This implies that $M_{k,k+1}$ has at
least one eigenvalue $\lambda$, such that
\begin{eqnarray}
\nonumber
|\lambda-\langle \Psi_a,M_{k,k+1} \Psi_a \rangle|
\\ \nonumber
\leq
J l_{proj} {\cal O}(\exp[-l_{proj}/(\log_2(N) l_{fac})]).
\end{eqnarray}
Therefore, there is at least one state $\alpha^2_k\rangle$ for each site $k$.

For each pair of sites $k,k+1$
surrounded by an oval on the second line of Fig.~2(b), we calculate the
$P^1_k$ of the above paragraph.  Then, for each group of four sites,
$k,k+1,k+2,k+3$ surrounded by any oval on the third line of Fig.~2(b),
we let $P^2_k$ project onto the eigenvectors of
$P^1_k P^1_{k+2} M_{k+1,k+2} P^1_{k+2} P^1_k$ in the
space of states $\alpha^2_k\rangle \otimes \alpha^2_{k+2} \rangle$, such
that the eigenvector has
eigenvalue
$\lambda$ such that 
\begin{eqnarray}
\nonumber
|\lambda-\langle \Psi_a, M_{k+1,k+2} \Psi_a \rangle|\\ \nonumber
\leq
J l_{proj} [{\cal O}(\exp(-l_{proj}/l_{fac}))+2 b_1]^{1/\log_2(N)}.
\end{eqnarray}
Let $\alpha^3_k$ label the resulting eigenvectors $\alpha^3_k\rangle$.
Define $\Psi^1_a=P^1_k P^1_{k+2} \Psi_a\rangle/
\sqrt{\langle \Psi_a,P^1_k P^1_{k+2} \Psi_a \rangle}$.
This vector is normalized to unit norm.  Then,
$|\Psi^1_a-\Psi_a|\leq 2 b_1$.
Thus, 
\begin{eqnarray}
\nonumber
|(M_{k+1,k+2}-\langle \Psi^1_a,M_{k+1,k+2} \Psi^1_a\rangle)\Psi^1_a\rangle|
\\ \nonumber\leq J l_{proj} {\cal O}(\exp(-l_{proj}/l_{fac}))+2 b_1 J l_{proj}.
\end{eqnarray}
Therefore,
$\langle \Psi^1_a, P^2_k \Psi^1_a \rangle \geq 1-b_2$,
where
$b_2=(\exp(-l_{proj}/l_{fac})+2 b_1)^{1-1/\log_2(N)}$.

Proceeding in this fashion, we find that
$\langle \Psi^{m-1}_a, P^m_k \Psi^{m-1}_a \rangle \geq 1-b_m$,
where
$b_m=(\exp(-l_{proj}/l_{fac})+2 b_{m-1})^{1-1/\log_2(N)}$.
There are $h\equiv\log_2(N)$ levels of this construction.
Thus, after the last step, $b_{h}$ is bounded by a quantity
of order $N \exp[-(l_{proj}/l_{fac})(1-1/h)^h]
\sim
N \exp(-l_{proj}/el_{fac})$.  Choose $l_{proj}$ such that
$l_{proj}/l_{fac}\geq {\cal O}(\log(J/\Delta E)\log(N)^2)$.  Then,
there is at least one state $\alpha^h\rangle$, and one may
show that
\begin{eqnarray}
\label{hchy}
\langle \alpha^{h}, \sum_i M_i \alpha^{h} \rangle\leq
\langle \Psi_a, \sum_i M_i \Psi_a \rangle+\\ \nonumber N
J {\cal O}(\exp[-l_{proj}/(\log_2(N) l_{fac})]).
\end{eqnarray}
Thus, the energy of this state $\alpha^{h}\rangle$ will be within $\Delta E$ of
the ground state energy.

Thus, this procedure yields a matrix product state close to the ground state in
energy.  At the same time, this procedure does not yield too many states.  All of
the states
$\alpha^{h} \rangle$ are within energy of order $\Delta E$ of the ground
state.  We can choose $l_{proj}$ so that they lie within energy $\Delta E/2$
of the ground state; it can then be shown that the number of distinct
$\alpha^{h}\rangle$ is at most $2n$, where $n$ is the number of ground
states.  By choosing the bound on the expectation value of the energy even
smaller, the ground state may be approximated to arbitrary accuracy.

\begin{figure}[tb]
\centerline{
\includegraphics[scale=0.7]{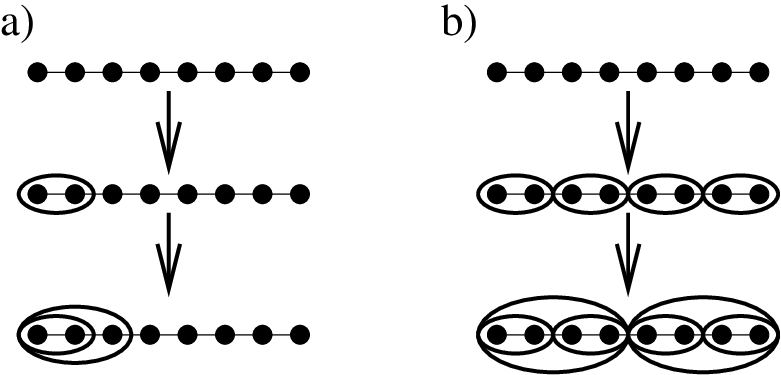}
}
\caption{Illustration of iterative and hierarchical procedures.
a)  Iterative procedure.  Filled circles represent sites, ovals surrounding
sites representing grouping of sites into a supersite.  The first
stage groups sites $1$ and $2$.  The second groups the combined site with
site $3$, and so on.
b)  Hierarchical procedure.  The first stage groups pairs of sites,
the second stages groups four sites into a single site, and so on.}
\end{figure}

This procedure
can be extended to systems in more than one
dimension.  One possibility is to
treat a higher
dimensional system as a one-dimensional system by grouping all the sites
with a particular value of a coordinate into a supersite as described in
the subsection reviewing the matrix product and valence bond constructions.
Another possibility is to group
group $2^d$ sites together at each stage of the hierarchy.
In either case, it is still only necessary to take
an $l_{proj}$ which grows as a power of $\log(N)$ to get a good
approximation to the ground state.

\section{Quantum Simulation}
One application to quantum simulation of matrix product or
valence bond states in higher dimensions is variational\cite{2d}, and
the results here may be useful in analyzing such algorithms.
However, another application of these results to quantum simulation involves
using the construction of the previous sections.
The results in this section are
illustrative, and will be worked out for more practical
examples in a future publication\cite{tbp}.
The goal here is to show, for at least some simple examples, that
in principle
one can actually use computations on finite systems to write down
wavefunctions for much larger, or even infinite, systems, and to provide
a variational technique that gives a {\it lower} bound on the energy.

To do this, we propose
to calculate the matrices $M_i$, determine the correct
eigenvalue of each $M_i$, and then use this to determine
the ground state wavefunction for a large system.
At this point, there is a very natural objection.  The procedure
requires that we find the correct eigenvalue of the $M_i$.  If
we can do this, then we know the ground state energy of the
system (up to an exponentially small error).  If $l_{proj}$ is sufficiently
large to obtain approximately correct eigenvalues for the $M_i$, then why not
just do an exact diagonalization of the system on a system of
size $l_{proj}$ and compute the ground state energy that way?  One
answer is that the ground state energy is not the only interesting
aspect of the system.  Correlation functions are much more important,
and it is often a difficult task to determine the long-range order
from the quantum numbers
(such as momentum and spin) of the low-lying states found in exact
diagonalization.  The procedure we outline of sewing
together solutions will provide a way of
taking a solution on a finite size system and extending it to
a wavefunction for a system of much larger, or even infinite,
size.  This wavefunction can then be used to compute long-distance
behavior of correlation functions.  Also, as we will see below, in some
cases this yields better energy estimates than exact diagonalization of the
same size system.

We now discuss the first step, computing the $M_i$.
One way is to  directly calculate the $M_i$, using
Eq.~(\ref{mid}) and using Eq.~(\ref{odef}) to define
$H_i^{\rm trunc}(t)$.  One way to compute
$H_i^{\rm trunc}(t)$ is via a series expansion:
$H_i^{\rm trunc}(t)=H_i+i t[H^{\rm loc},H_i]+...$
Another way is to exactly diagonalize a finite system of size
$l_{proj}$, compute the matrix elements of
$H_i$ between the eigenvalues, and use Eq.~(\ref{mel}) to get
matrix elements of $\tilde H_i^0$.  In both cases
it is necessary to choose $q$ to minimize
the difference $||\tilde H_i^0-M_i||$ given by Eq.~(\ref{apb}) and also the
norm $|P_{high} \tilde H_i^0\Psi_a\rangle|$ of Eq.~(\ref{gs}), giving a $q$ of
order $c_1 l_{proj}\Delta E $.

Given that the ground state is an approximate eigenstate of the
$M_i$ computed in this way, the question arises: what is the eigenvalue?
That is, what is
the approximate expectation value of $\langle H_i \rangle$ in the
ground state?  
In some cases, finding the correct
eigenvalue is an extremely difficult problem.
For example, consider an Ising spin-glass Hamiltonian: $H=\sum_i H_i$
with $H_i=\sum_j J_{ij} S^z_i S^z_j$, where $J_{ij}$ is some set
of random couplings between nearby spins.  This is a purely classical
problem, since all of the $H_i$ commute, and any state in which each spin
has a definite value of $S^z$ is an eigenvector of every $H_i$.  However,
to find the ground state is clearly a difficult task!  In this highly
disordered system, each $H_i$ has a different expectation value and one
must find the correct eigenvalue for each one.  For ordered systems the
task is much easier.  If the ground state does not break translational
symmetry then each $M_i$ has the same ground state expectation value.
If there is a symmetry breaking ground state with an enlarged unit cell,
there are still only a discrete number of different expectation values
for the $M_i$.  For example, in the Majumdar-Ghosh chain, the ground
states are invariant under translation by two sites, and there are
two different ground state expectation values for the $M_i$.

We have performed some simple numerical experiments on systems of free
particles on a lattice with different gapped band structures.
In this case there is no symmetry breaking, and
if $H$ is translationally invariant, then the expectation values of the
$M_i$ in the ground state are also translationally invariant.
The lowest eigenvalue of $\tilde H_i^0$ for given $q$ always
provides a lower bound to
the energy as in Eq.~(\ref{varnum}).
However, we have found that as $q$ is increased, the lowest
eigenvalue increased and
converges rapidly to the ground state energy per site.  There
are several other eigenvalues very close in energy to the lowest, and then
a gap to the rest, so in this case at least the identification of the
correct eigenvalue is easy and this technique in fact provides a way to
compute the energy per site.  An upper bound to the energy per site as
well as a macroscopic wavefunction can
be obtained by using one of the matrix product constructions above.  The
most important question, of course, is how well this procedure can be extended
to complicated interacting systems.

Although the main point of the present paper is the formal construction
of the $M_i$, we finally
discuss here two brief attempts to apply these techniques to interacting
systems.  
First, 
the techniques here provide a variational lower bound to the
energy of the system.  Consider a translationally invariant Hamiltonian,
$H=\sum_i H_i$, with all the $H_i$ equal.  Then, the ground state
energy per site, $E_0/N$ is at least equal to the smallest eigenvalue of
$H_i$.  Further,
\be
\label{varnum}
E_0/N\geq \lambda_{\rm min}(H_i+i a_1 [H_i,H] + a_2 [[H_i,H],H] + ...),
\ee
where $\lambda_{\rm min}(O)$ is equal to the smallest eigenvalue of $O$,
and $a_1,a_2,...$ are arbitrary constants.  The operators $\tilde H_i^0$
are given by a particular choice of the constants $a_k$ that can be obtained
from Eq.~(\ref{tnd}).  Other choice are possible, and varying over
the constants will provide a variational lower bound for the energy.  Thus, this
provides an interesting complementary approach to other
quantum simulation techniques, since almost all other techniques provide
either approximate estimates or variational {\it upper} bounds.

We have studied how this bound is approached on a spin-$1$ Heisenberg
chain.  The ground state energy of this chain is known very accurately
from DMRG.  The present method (\ref{varnum})
here is not intended to compare to DMRG,
but rather we compare to exact diagonalization.  In essence, this
method provides another type of boundary condition, instead of the usual
periodic or anti-periodic boundary condition, with the advantage that
in this case we know rigorously how the ground state energy compares to
the energy from this procedure.  We have some freedom to pick the constants
$a_1,a_2,...$.  We also have freedom to choose $H_i$ to be an operator
on a pair of neighboring sites, on three neighboring sites, or in general
on any supercell of $m$ sites.  Since $H$ is real symmetry
in this case, we pick $a_k=0$ for $k$ odd.  If $m=2$, at the most trivial level
of $a_k=0$ for all $k$, we need to diagonalize a Hamiltonian with
$n=2$ sites.
For $a_2\neq 0, a_k=0$ for $k>2$, we need to diagonalize a Hamiltonian
with $n=6$ sites, and so on.  If we instead we pick $m=3$ we need to
diagonalize a Hamiltonian with $n=3$ sites if all $a_k=0$, a
Hamiltonian with $n=7$ sites
if $a_2\neq 0, a_k=0$ for $k>2$, and a Hamiltonian with $n=11$ sites
if $a_2\neq 0, a_4 \neq 0, a_k=0$ for $k>2$.

We have chosen to take $m$ odd
for the following reason.  This technique provides
a lower bound for the energy of the Hamiltonian, both for the infinite system
and for the particular $n$-site Hamiltonian.  However, we know for the
spin-$1$ chain that on even size systems the energy obtained is already less
than the ground state energy.  Thus, the most efficient results will be obtained
for odd size systems, and hence we pick an odd $m$.

For $m=3$ and $a_k=0$ for all $k$ we find $E_0/N\geq -1.5$, which is equal
to one-half the energy of an open three site chain.  For $m=3$ and $a_2\neq 0$,
with all higher $a_k=0$, we find $E_0/N\geq -1.42569$ after picking
$a_2=-0.075$.  This requires diagonalizing a 7 site system; the most
naive estimate that can be obtained from a 7 site system, taking $m=7$ gives
a worse bound of $E_0/N\geq -1.43909$.
For $m=5, a_2=-0.075$ and all higher $a_k=0$, we find
$E_0/N\geq -1.4156$.  Slight improvements on these bounds
can be found by better optimization of $a_2$.
This last estimate requires diagonalizing a 9 site system.
One we move to an 11 site system, we have the option of either considering
$m=7, a_2 \neq 0$, or $m=5, a_2 \neq 0, a_4 \neq 0$, with all higher $a_k$
vanishing.

For purposes of computing the energy, then, this technique offers some
slight improvements over exact diagonalization.  Using exact diagonalization
with periodic boundary conditions
in this particular case, even size systems offer lower bounds, while odd size
systems offer upper bounds, while we find a lower bound in every case.
The estimate from a 9 site system using this method, for example, is better
than that found using exact diagonalization of an 8 site system, where
the energy is estimated to be $-1.417$, but not as good as that found by
diagonalizing a 10 site system, where one finds $-1.4094$, while for
a 7 site system the estimate is better than that found from diagonalizing
a six site system, where one finds $-1.44$.  Further this method offers
the only way to obtain rigorous lower bounds.  This may become especially
important in studying frustrated systems.  On a frustrated system with
spiral order, for example, one has no foreknowledge that exact diagonalization
of a periodic chain of a given length will provide a lower or upper estimate
on the energy.  The energy estimates of this unfrustrated chain obtained
from exact diagonalization of odd size systems are poor compared
to those obtained by exact diagonalization of even size systems;
on a frustrated chain one has no notion of which sizes will yield accurate
estimates of the energy.

In one dimension on unfrustrated systems, then, this technique does
not give much improvement on the energy, as one can obtain a better
estimate by exact diagonalization of a system of one site more.
However,
in higher dimensions, even on an unfrustrated system, this technique
may become more useful again.  Suppose we have an unfrustrated system
of size $L$-by-$L$ in higher dimensions, and suppose it follows the pattern
found here, that the most accurate estimate of the energy from exact
diagonalization is found from periodic systems with even $L$ where one obtains
a lower bound.  The present technique offers the possibility of obtaining
accurate estimates of the energy from a system of size $L-1$ instead of
size $L$, which means studying a system with $2L-1$ fewer sites.

We know
that the ground state wavefunction of the full system has a bounded
projection onto states other than those with close to the given eigenvalue of
$H_i+a_2[[H_i,H],H]+...$ .  Indeed, for the particular choice of
Eq.~(\ref{tnd}), the projection onto such states is provably exponentially
small in the size of the system considered.  Thus, one can follow
a procedure of breaking a chain or lattice into blocks, building the
$M_i$ for each block and diagonalizing it in each block,
restricting to the states with the given eigenvalues in each block,
and then studying the behavior of the full Hamiltonian in this reduced
space of states.  
As a first test of this algorithm, we take $m=3$.  If we simply take the
project onto the states with lowest
eigenvalues of the three site Hamiltonian, all $a_k=0$, this provides
a poor approximation to the eigenstates of larger systems.  For example, on
a seven site chain, with the three sites taken in the middle of the
chain, there are $3^5=243$ states such that the three site Hamiltonian has
an energy per bond given by
$E/2=-1.5$.  The Hamiltonian of the seven site chain with open
boundary conditions has a lowest energy state with energy per bond equal to
$-1.43909$, but if we project onto the given $243$ states above, we only
achieve an energy per bond of $-1.36496$.
However, if we take
$a_2=-.075$ with $a_k=0$
for $k>2$, there are $243$ eigenvalues of the Hamiltonian with $E/2\leq -1.34$,
and then the $244$-th state has eigenvalue $-0.945706$.  Projecting the
Hamiltonian of the seven site chain with open boundary onto
these $243$ states we find that an energy per bond of $-1.43331$.  Thus,
we have accurately selected the needed states, having projected from
$2187$ states to $243$ states, or from
$393$ states with $S_z=0$ to $51$ states with
$S_z=0$.

A further test breaking a fourteen site periodic
spin-1 chain into
seven site blocks and taking $m=3$ and $a_2=-.075$ to project onto states
in each block showed that the lowest energy wavefunction in this subspace
had energy per site equal to $-1.39545$, compared to the exact
result of $-1.40394$ for this size.
If instead of joining subblocks
we had added sites to a subblock one at a time, and
used the present method of projecting onto the states in each subblock, we would
arrive at an algorithm similar to DMRG.  However, the goal is
not to compare to DMRG in
one dimension, but rather to present an algorithm that
can be extended to higher dimensions, which DMRG
cannot.

A second technique, which may also offer the possibility of improved accuracy
on the energies compared to exact diagonalization is based on the idea that
the operators $\tilde H_i^0$ are defined by the choice of constants
$a_k=0$ for $k$ odd, and $a_k=(-1)^{k/2}[(k-1)!!/k!] 
(q/\Delta E^2)^k$ for $k$ even.
We then have
a series expansion for $\tilde H_i^0$ in powers of $q$.  We can then
perturbatively expand the eigenvalues of $\tilde H_i^0$ in
powers of $q$ and extrapolate
to $q=\infty$ from a finite number of terms.  This is a speculative approach
that is currently being studied.

\section{Non-zero Temperature}
We now turn to systems at non-zero temperature.  In this case the
temperature enables us to construct an approximate matrix product
form for the density matrix, regardless of whether or not there is
a gap.  The unnormalized density matrix for the system is equal to
$\rho=\exp(-\beta H)$.

We will construct an approximate matrix product form,
$\rho(\beta,l_{proj})$, so that
\begin{eqnarray}
\label{mpf}
\rho(\beta) \approx 
\rho(\beta,l_{proj})
\\ \nonumber
=\sum_{\{\alpha_k\}}
\rho_1(\alpha_1) \rho_2(\alpha_2) ...
F_1(\{\alpha_j\}) F_2(\{\alpha_j\}) ...
\end{eqnarray}
where for each site $i$ we assign an index $\alpha_i$ defined below,
and sum over all values of that index.  The operator $\rho_i(\alpha_i)$
acts only on site $i$, and the functions $F_i$ obey a finite range
constraint: each $F_i$ depends only on the $\alpha_j$
for $d(i,j)\leq l_{proj}$ for some $l_{proj}$ defined below.
The error between $\rho(\beta)$ and $\rho(\beta,l_{proj})$ will be exponentially
small in $l_{proj}$, while the range of the indices $\alpha_i$ will depend
on $l_{proj}$.
Specifically, we bound the error by showing that
for any operator $O$,
$Z^{-1} |{\rm Tr}[O \rho(\beta)]-{\rm Tr}[O \rho(\beta,l_{proj})]| \leq
c ||O||$, for some constant $c$, where we define $Z={\rm Tr}(\rho(\beta))$.

Before defining $\rho(\beta,l_{proj})$, we recall the
Trotter-Suzuki\cite{trotter} decomposition
of the path integral.  In this case, we write $\rho(\beta)\approx
\rho_n$, where $\rho_n=[\prod_i \exp(-\beta H_i/n)]^n$, where
the product ranges over all $i$ in some given sequence; since the
different $H_i$ do not commute, the result depends on the particular
sequence chosen.  We claim that
each of the $\rho_n$ can be written exactly in a matrix product form as
in the right-hand side of Eq.~(\ref{mpf}).
A given operator $\exp(-\beta H_i/n)$ acts on sites within a distance
$R$ of site $i$, and can be written as $\exp(-\beta H_i/n)=
\sum_{\{\alpha_{i,j}\}}
F_i(\{\alpha_{i,j}\}) \prod_{j,d(i,j)\leq R} O_j(\alpha_{i,j})$, where
the operator $O_j$ acts only on site $j$ and the
range of values of index $\alpha_{i,j}$ is exponentially large in $S$,
the number of sites within distance $R$ of site $i$.
Here, $F_i(\{\alpha_{i,j}\})$ is
some function of the $S$ different indices $\alpha_{i,j}$ with
the given $i$.
The operator $\rho_n$ is a finite product of these operators
$\exp(-\beta H_i/n)$.  Each term in this product can be written
in matrix product form.  For each site $i$, the operator
$\exp(-\beta H_i/n)$ appears $n$ times in the given product, and on
the $m$-th time it appears, we use a set of indices $\alpha_{i,j}^m$ to provide
the matrix product form as above.
For each site, $j$, we have $n$
different indices $\alpha_{i,j}^m$ for each site $i$ within distance $R$ of
site $j$, and thus at most $nS$ different indices for each site
$j$.  Grouping all such indices for a given site $j$ into
one index $\alpha_j$, and defining $F_i(\{\alpha\})=\prod_{m=1}^n
F_i(\{\alpha_{i,j}^m\})$,
we arrive at the matrix product form for $\rho_n$.

Thus, the Trotter decomposition gives an approximate matrix form, but the
error in this approximation compared to the exact result is not very good.
In contrast, the stochastic series expansion\cite{sse} provides a much better
way of approximating the desired $\rho(\beta)$ with much smaller error, but
it is difficult to write the stochastic series expansion result in a matrix
product form.
Below, we will propose a matrix product form with a bound on error comparable to
that in the stochastic series expansion.

\subsection{Percolation Transition at High Temperature}
The exponential $\exp(-\beta H)$ can be expanded as a power
series $1-\beta H + ...$ We will show that at sufficiently high temperatures,
$\beta^{-1}\sim J$, there is a ``percolation transition" in this exponential,
as we now describe.  Any given term in the power series expansion is a product
of $H_i$ for different sites $i$.  For each 
term, we define a set of ``active bonds" and ``clusters" as follows:
for each $H_i$ which appears in the given term, we connect by active bonds
all sites acted on by $H_i$,
so that
the length of the active bonds is at most $2R$.  We define a cluster to be
a set of sites, all connected to each other by active bonds, and not connected
to any other sites outside the cluster by active bonds.
Then, for $\beta\leq \beta_0$, where $\beta_0$ is specified below,
define $\rho(\beta,l_{proj})$
to include only the terms in the power series
such that no two sites $i,j$, with $d(i,j)>l_{proj}$,
are in the same cluster.
Thus, each term in $\rho(\beta,l_{proj})$ is a product of
operators, each operator acting on the sites within a given
cluster, such that each cluster has a diameter at most $l_{proj}$.
See Fig.~2.  For $\beta\leq\beta_0$, at temperatures
above the percolation transition, we will
be able to bound the difference between $\rho(\beta)$ and 
$\rho(\beta,l_{proj})$.

\begin{figure}[tb]
\centerline{
\includegraphics[scale=0.7]{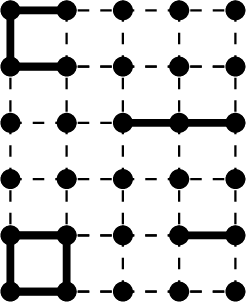}
}
\caption{Example of a set of active bonds shown as solid
lines; dashed lines represent bonds which are not active while
circles represent sites.
The Hamiltonian is a sum of terms
which act on pairs of neighboring
sites, so that the bonds connect only neighboring
sites.  Using a Manhattan metric for the lattice, the set of active bonds
shown here is a term in $\rho(\beta,l_{proj}=2)$.  There are four
distinct clusters, three with diameter two and one with diameter one.}
\end{figure}

Let $C$ be some set of sites $i$.  Define $H_C=\sum_{i\in C} H_i$.
Define $B(C)$ to be equal to the set of all sites $j$  such that
$j\neq i$ for any $i\in C$ and such that $H_j$ acts on a site $i$ for
some site $i\in C$.  Then, $H_{B(C)}=\sum_{i\in B(C)} H_i$.
Then, if a given term in the power series expansion
for $\exp(-\beta H)$ includes $H_i$ for every $i\in C$, but does
not include $H_j$ for all $j\in B(C)$,
then this term in the power series expansion has a cluster which includes
exactly the sites acted on by $H_i$ for all $i\in C$.

For any operators 
$O_1,O_2,...$, 
\begin{eqnarray}
\nonumber
Z^{-1} {\rm Tr}(O O_1(i\tau_1)
O_2(i\tau_2)...
 \exp[-\beta H])\leq
||O|| ||O_1|| ||O_2|| ...,
\end{eqnarray}
where $O(i\tau)=\exp(-H \tau) O \exp(H \tau)$
and $0\leq \tau_1 \leq \tau_2  \leq ... \leq \beta$.
Thus, if $C$ has $n_C$ sites and $B(C)$ has $n_{B(C)}$ sites, we find that
for any operator $O$,
\begin{eqnarray}
\label{ddb}
Z^{-1} {\rm Tr}(O \exp[-\beta (H-H_C-H_{B(C)})])
\\ \nonumber
\leq
||O||
\exp[(n_C+n_{B(C)})J\beta],
\end{eqnarray}
as can be seen by using a power series expansion of
\begin{eqnarray}
\nonumber
Z^{-1} {\rm Tr}(O \exp[-\beta (H-H_C-H_{B(C)})])
\\ \nonumber
=
Z^{-1} {\rm Tr}(O
{\cal T} \exp[\int_{0}^{\beta} {\rm d}\tau (H_C(i\tau)+H_{B(C)}(i\tau))]
\exp[-\beta H]),
\end{eqnarray}
where ${\cal T}$ gives the $\tau$-ordered exponential.

We now define $\rho(\beta,C)$ to be equal to the sum of all terms
in the power series expansion  of $\exp(-\beta H)$ that include $H_i$ for
all $i\in C$, but do not include $H_j$ for any $j\in B(C)$.
For any $O$, 
\begin{eqnarray}
\nonumber
{\rm Tr}(O \rho(\beta,C))
\leq [\exp(J \beta)-1]^{n_C}\times \\ \nonumber\exp[\beta ||H_C+H_{B(C)}||]
{\rm Tr}(O \exp[-\beta H],
\end{eqnarray}
as can be
seen by a power series expansion.  
Using $||H_C+H_{B(C)}||\leq (n_C+n_{B(C)}) J \leq S J n_C$ gives
\begin{eqnarray}
\label{percb2}
Z^{-1} {\rm Tr}[O \rho(\beta,C)]
\\ \nonumber
\leq
||O|| 
[\exp(J \beta)-1]^{n_C}
\exp[n_C S J\beta].
\end{eqnarray}

Then, $\rho(\beta)-\rho(\beta,l_{proj})$
is a sum over terms which include a sequence of
active bonds connecting any two sites $i,j$ separated by a distance at least
$l_{proj}$.  This difference
$\rho(\beta)-\rho(\beta,l_{proj})=-\sum_{m=1}^{\infty} (-1)^m \rho_m$,
where $\rho_m$
is equal to the sum of all terms in the power series expansion of $\rho$
which include at least $m$ clusters with diameter greater than $l_{proj}$. 
Using Eq.~(\ref{percb2}):
$Z^{-1} {\rm Tr}(O \rho_1)\leq
\sum_{\rm clusters} y^{n_C} ||O||$,
where the sum ranges over all different clusters on the lattice which connect
two sites separated by distance greater than $l_{proj}$ and $n_C$ is the number
of sites in the cluster and where $y=
[\exp(J \beta)-1]
\exp[S J\beta]$.
Similarly,
$Z^{-1} {\rm Tr}(O \rho_m)\leq
(1/m!) (\sum_{\rm clusters} y^{n_C})^m
||O||$.

{\bf Remark added:} In response to some e-mail questions which made me realize that the above paragraph was unclear, here is some additional explanation in this paragraph and the next two paragraphs.  We have defined $\rho(\beta,C)$ for a given cluster $C$ above.  In the same way as we have defined $\rho(\beta,C)$, we can also define $\rho(\beta,C_1,C_2)$ in which both $C_1,C_2$ are clusters of active bonds, and similarly we can define $\rho(\beta,C_1,C_2,C_3)$, and so on.  In the same way as we bounded the trace of $\rho(\beta,C)$ with an arbitrary operator $O$ (Eq.~(\ref{percb2})), we can derive similar bounds on $\rho(\beta,C_1,C_2)$ which would bound its trace with arbitrary operators by even smaller quantities, and so on.    Note, by definition of a cluster, the clusters $C_1,C_2$ in $\rho(\beta,C_1,C_2)$ have to be disjoint.

Now consider the term $\rho(\beta,C)$.
 This term $\rho(\beta,C)$ is a sum of many different terms in the perturbation series.  If cluster $C$ has diameter greater than $l_{proj}$, then every term in the perturbation series for $\rho(\beta,C)$ has at least one cluster with diameter bigger than $l_{proj}$, namely the cluster $C$ itself.  However, some of the terms that contribute to the perturbation series for $\rho(\beta,C)$ may have more than one cluster with diameter bigger than $l_{proj}$: a given term might happen to include some other large cluster as well as $C$.  Let us define $\rho_1$ to be the sum of $\rho(\beta,C)$ over clusters $C$ with diameter larger than $l_{proj}$.  Define $\rho_2$ to be the sum of $\rho(\beta,C_1,C_2)$ over clusters $C_1,C_2$ with diameter larger than $l_{proj}$, define $\rho_3$ to be the sum of $\rho(\beta,C_1,C_2,C_3)$ over clusters $C_1,C_2,C_3$ with diameter large than $l_{proj}$, and so on.

Every term in $\rho_1$ has at least $1$ cluster with diameter larger than $l_{proj}$ and in general every term in $\rho_m$ has at least $m$ clusters with diameter larger than $l_{proj}$ (this what is meant by ``which include at least $m$ clusters" above).  Note that $\rho-\rho_1$ does not give us the desired result of summing all terms in the perturbation series
 which do not include any clusters with diameter larger than $l_{proj}$ as any term in the perturbation series with two clusters with diameter larger than $l_{proj}$ will be subtracted off twice; we will have overcounted.  However, $\rho-\rho_1+\rho_2-\rho_3+...=\rho+\sum_{m=1}^{\infty} (-1)^m \rho_m$ gives the desired result.  Everything that is overcounted by subtracting off too many times in $\rho_1$ is added back on in other terms; this is an inclusion-exclusion sum.

We now show the ``percolation transition".
The number of clusters of $n_C$ sites is bounded on a regular
lattice by $Nx^{n_C}$ for some constant $x$, as is known for
lattice animals\cite{klarner}, where $N$ is the number of sites on the lattice.
Thus for sufficiently small $\beta$, 
\begin{eqnarray}
\nonumber
\sum_{m=1}^{\infty} (1/m!) (\sum_{\rm clusters} y^{n_C})^m\leq
\exp[N \sum_{n=l_{proj}+1}^{\infty} (xy)^n]-1.
\end{eqnarray}
For fixed $J, S$, we
can make $y$ as close to zero as desired by taking sufficiently small
$\beta$.  Thus, there exists some small enough $\beta_0$ such that
for $\beta\leq\beta_0$,
\begin{eqnarray}
\nonumber
\sum_{m=1}^{\infty} (1/m!) (\sum_{\rm clusters} y^{n_C})^m\leq 
\exp[N \exp(-l_{proj}/\xi_{perc})]-1,
\end{eqnarray}
where
$\xi_{perc}$ is of order the interaction range $R$.
Then,
\begin{eqnarray}
\label{perc2}
Z^{-1}|{\rm Tr}[O \rho(\beta)]-{\rm Tr}[O \rho(\beta,l_{proj})]|\\ \nonumber
\leq
\{\exp[N \exp(-l_{proj}/\xi_{perc})]-1\}  ||O||.
\end{eqnarray}

\subsection{Matrix Product Form at High Temperature}
We now show that $\rho(\beta,l_{proj})$ realizes the matrix product form
Eq.~(\ref{mpf}) for $\beta\leq\beta_0$.  Each term in
$\rho(\beta,l_{proj})$ is a product of terms acting on clusters of sites.
We now
specify the indices $\alpha_i$.  First, if site $i$ is in a cluster, the
index
$\alpha_i$ indicates all the sites in that cluster.
There is some redundancy in this
description: if two sites $i$ and $j$ are in the same cluster, then the
indices $\alpha_i$ and $\alpha_j$ specify the same set of sites.
Each $\alpha_i$ will also specify some additional
information.  The sum of all terms which include a given cluster is some
operator acting on the sites in the cluster; this operator can be decomposed
into a sum of products of operators that act on individual sites in the
cluster.  The indices $\alpha_i$ will also keep track of the terms in this
sum.
The number of different values that each index $\alpha_i$ can assume
is exponentially large in $l_{proj}^d$ in this construction.  However, this
is much smaller than the number of different values that would be needed
to describe a general operator $\rho$: in general, one might need a range
of values which is exponentially large in the system size.
Thus, $\rho(\beta,l_{proj})$ can be exactly written in a matrix product form,
as in the right-hand side of Eq.~(\ref{mpf}).  

Further, we have shown that
$\rho(\beta,l_{proj})$ is a good approximation to $\rho(\beta)$.
This can be expressed in terms of the trace norm.
For any operator $O$, the trace norm $|O|$
is equal to ${\rm Tr}(\sqrt{O^{\dagger} O})$, where ${\rm Tr}(...)$ denotes the
trace over all states and the unique positive square-root is taken.  For
a Hermitian operator, such as $\rho(\beta)$ or $\rho(\beta,l_{proj})$,
$|O|$ is equal to the sum of the absolute
values of the eigenvalues.  For a positive definite Hermitian operator,
such as $\rho(\beta)$, the trace norm is equal to the trace.
Use Eq.~(\ref{perc2}), valid for arbitrary $O$, and choose $O$ so that
in a basis of eigenvectors of $\rho(\beta)-\rho(\beta,l_{proj})$, $O$
is a diagonal matrix with each diagonal entry equal to $\pm 1$, the sign
being chosen the same as the sign of the corresponding eigenvalue.
Then,
\be
|\rho(\beta)-\rho(\beta,l_{proj}) |\leq \{\exp[N \exp(-l_{proj}/\xi_{perc})]-1\}
|\rho(\beta)|.
\ee
By taking an $l_{proj}$ that is of order $\log(N) \xi_{perc}$, we
can obtain a small error $|\rho(\beta)-\rho(\beta,l_{proj})|$.

We now obtain a matrix product form at arbitrary temperature
$\beta$ that guarantees positive definiteness of $\rho(\beta,l_{proj})$.
For $\beta>\beta_0$,
set 
\be
\label{mplt}
\rho(\beta,l_{proj})=\rho(\beta/n,l_{proj})^n,
\ee
where
$n$ is the smallest even integer such that $\beta/n\leq \beta_0$.
By picking $n$ even we guarantee that $\rho(\beta,l_{proj})$ is positive
definite.  Eq.~(\ref{mplt}) realizes a matrix product form (\ref{mpf}):
for each site $i$ there are $n$ different indices $\alpha_i$, one from
each of the $\rho(\beta/n,l_{proj})$.  We group these into one single index,
and call the new index $\alpha_i$, giving Eq.~(\ref{mpf}).
The number of different values that the new $\alpha_i$ can assume
is exponentially large in $(\beta/\beta_0) l_{proj}^d$.  We can bound
the error at arbitrary temperature as follows.

For each $\rho(\beta/n,l_{proj})$ we define clusters.  We can
show, as above, that
\begin{eqnarray}
\nonumber
Z^{-1}{\rm Tr}(O\rho(k\beta/n)\rho(\beta/n,C)\rho((n-k-1)\beta/n))
\leq ||O|| y^{n_C},
\end{eqnarray}
for any $0\leq k\leq n-1$,
where
$y=[\exp(J \beta/n)-1]
\exp[S J\beta/n]$.

Define $\rho_m$ to be the sum over all terms in which there are
at least $m$ clusters with diameter greater than $l_{proj}$.
Then, 
\begin{eqnarray}
\nonumber
Z^{-1}{\rm Tr}(O\rho_m)\leq (1/m!)(\sum_{clusters} n y^{n_C})^m
||O||.
\end{eqnarray}
Following the same arguments as before
\begin{eqnarray}
Z^{-1}|{\rm Tr}[O \rho(\beta)]-{\rm Tr}[O \rho(\beta,l_{proj})]|\\ \nonumber
\leq
\{\exp[(\beta/\beta_0) N \exp(-l_{proj}/\xi_{perc})]-1\}  ||O||.
\end{eqnarray}
and so we need an $l_{proj}$ that grows logarithmically in the
system size to obtain a good approximation.
At the value of $l_{proj}$ that gives a good approximation,
$l_{proj}\sim\sum \xi_{perc} \log(N)$, the number of different values
that each index $\alpha$ can assume is of order
\be
\exp[{\cal O}(\xi_{perc}^d \log(N)^d(\beta/\beta_0)].
\ee

The final matrix product form turns a quantum statistical mechanics
problem into a classical statistical mechanics problem, as tracing over
the $\rho_i$ gives a probability distribution for the $\alpha_i$.
However, the resulting classical problem may suffer from a sign problem.

\section{Discussion}
We have shown how to construct an approximate local projective
form of short-range, gapped Hamiltonians, and used this to
show that the ground states are close to matrix product states.
The main goal of the present paper is at a formal level; the techniques
developed in this paper provide a way to, in principle at least, extend
calculations on small systems to wavefunctions on much larger system sizes.
It is worth comparing to \cite{mudry}, which showed how to write
a very general class of Hamiltonians as a sum of projection operators.
The projection operators there were two-by-two matrices which makes
it much easier to find the ground state.  Here, the projection
operators are large matrices, and the task of constructing
the ground state from these operators is tricky.  However, the important
advance here is that the projection operators are local.  This
strongly constrains the ground state and leads to the matrix product
for the ground state.

In addition to the formal interest,
this work may find practical use in numerical simulation, as,
at least in simple cases, it is possible to directly calculate
the local projective form.  Further, we have provided (\ref{varnum}), a
variational lower bound on energy.

We finally return to the impurity problem raised in the introduction.  The
following procedure will generate a good wavefunction for the whole system.
First, numerically compute the $M_i$ for some region around the impurity.
Next, compute the appropriate $M_i$ for the 
system outside that region; this can be done analytically since that part of the
system is non-interacting.  Then, determine the appropriate eigenvalues
for the $M_i$ and get a basis of states
$\alpha_{impurity}\rangle\otimes\alpha_{insulator}\rangle$.  Finally,
use the $M_i$ that connect the two regions to determine a wavefunction
in this basis of states.  This procedure can be followed even if there
are many impurities embedded in the system and the resulting wavefunction
can be used as a starting point for further improvement.

{\it Acknowledgments---}
This work was supported by DOE contract W-7405-ENG-36.  

\appendix
\section{Locality}
The locality of the $\tilde H_i^0$
relies on the finite group velocity result\cite{fgv,lsmh,nchtgsim}.
Given the finite-range conditions on the Hamiltonian above, one can
bound the commutator
$||[A(t),B(0)]||$, where $A(t)=\exp(i{\cal H}t)A\exp(-i{\cal H}t)$, and
show that this commutator is exponentially small for times $t$ less than
$c_1 l$ where $l$ is the distance between $A$ and $B$ and $c_1$ is
some characteristic inverse velocity.
The bound is that
$||[A(t),B(0)|| \leq ||A|| ||B|| \sum_j g(t,d(A,j))$,
where the sum ranges over sites $j$ which
appear in operator $B$ and where the function $g$ has the property that
for $|t|\leq c_1 l$,
$g(c_1 l,l)$ is exponentially decaying in $l$ for large $l$ with
decay length $\xi_C$.  Also, for $t<t'$, $g(t,l)\leq (t/t') g(t',l)$.

Consider $[\tilde H_i^0,O_j]$ for some $O_j$ which
acts only on site $j$.
This equals
$(\Delta E/\sqrt{2 \pi q}) \int_{-\infty}^{\infty} {\rm d}t
[H_i(t),O_j] \exp[-(t\Delta E)^2/(2 q)]$.  Applying a triangle
inequality to the integral we have
\begin{eqnarray}
\nonumber
||[\tilde H_i^0,O_j]||\leq
(\Delta E/\sqrt{2 \pi q}) 
\int_{-\infty}^{\infty} {\rm d}t
||[H_i(t),O_j]]|| 
\times \\ \nonumber
\exp[-(t\Delta E)^2/(2 q)].
\end{eqnarray}
Let $l=d(H_i,j)$
be the distance between $H_i$ and site $j$.  We split the
integral over times $t$ into a sum of one integral over $|t|\leq c_1 l$
and one integral over $|t| \geq c_1 l$.
For $|t|\leq c_1 l$, we use the finite group velocity bound to
bound $||[H_i(t),O_j]||$ while
for $|t|\geq c_1 l$ we use $||[H_i(t),O_j]||\leq 2 J ||O_j||$.
Thus
\begin{eqnarray}
||[\tilde H_i^0,O_j]|| \leq \\ \nonumber
J ||O_j|| 
\Bigl(
\frac{\Delta E}{\sqrt{2 \pi q}}
\int_{|t|\leq c_1 l} g(c_1 l,l) \exp[-(t \Delta E)^2/(2 q)] 
+ \\ \nonumber
2 \frac{\Delta E}{\sqrt{2 \pi q}}
\int_{|t|\geq c_1 l} \exp[-(t \Delta E)^2/(2 q)]
\Bigr)  \leq
\\ \nonumber
J ||O_j|| (g(c_1 l,l) +
2 \exp[-(c_1 l \Delta E)^2/(2 q)]),
\end{eqnarray}
giving the claimed bound on the commutator (\ref{lbnd}).

We now bound the difference $||\tilde H_i^0-M_i||$.  We have
\begin{eqnarray}
\label{dmi}
||\tilde H_i^0-M_i|| \leq (\Delta E/\sqrt{2 \pi q}) \times
\\ \nonumber
\int {\rm d}t 
\exp[-(t \Delta E)^2/(2 q)] ||H_i(t)-H_i^{\rm trunc}(t)||.
\end{eqnarray}
The difference between $H_i(t)$ and $H_i^{\rm trunc}(t)$ is due to the
different Hamiltonians used to define the time evolution.

We now bound the difference between $H_i(t)$ and $H_i^{\rm trunc}(t)$.
This result will also be needed in the non-zero temperature calculation.
We can replace Eq.~(\ref{hpdef}) by
$H_{\rm loc}=\sum_{j,d(i,j)\leq l_{proj}-R}H_j+
\sum_{j,d(i,j)>l_{proj}+R} H_j$; by adding the terms
$H_j$ with $d(i,j)>l_{proj}+R$ we do not change $H_i^{\rm trunc}(t)$.
Then, 
\begin{eqnarray}
\nonumber
||H_i(t)-H_i^{\rm trunc}(t)|| 
\\ \nonumber
\leq
\sum_{j,l_{proj}-R<d(i,j)\leq l_{proj}+R}
\int_0^t {\rm d}t' ||[H_j,H_i(t)]||.
\end{eqnarray}
Using the finite group velocity bound we have
\begin{eqnarray}
\nonumber
||H_i(t)-H_i^{\rm trunc}(t) \\ \nonumber
\leq
\sum_{j,l_{proj}-R<d(i,j \leq l_{proj}+R}
J^2 S \int_0^t {\rm d}t' g(t',l_{proj}-2R),
\end{eqnarray}
where $l_{proj}-2R$ is the minimum distance between the operators
$H_j$ and $H_i$ and $S$ is defined to be the number of sites acted
on by the operator $H_j$.  However, the integral
$J S \int_0^t {\rm d}t' g(t',l_{proj}-2 R)$ is bounded by
$g(t,l_{proj})$\cite{lsmh}.  Thus,
\begin{eqnarray}
\label{thib}
||H_i(t)-H_i^{\rm trunc}(t)||\leq \sum_{j,l_{proj}-R<d(i,j)\leq l_{proj}+R}
J g(t,l_{proj})\\ \nonumber
\equiv N(l_{proj}) J g(t,l_{proj}),
\end{eqnarray}
where $N(l_{proj})$ is
defined to be the number the number of sites $j$ with
$l_{proj}-R<d(i,j)\leq l_{proj}+R$.

Eq.~(\ref{thib}) can be expressed in a more general form.
For any operator $O_i$, we can define $O_i^{\rm trunc}(t)$ by
$O_i^{\rm trunc}=\exp(i H_{\rm loc} t) O_i \exp(-i H_{\rm loc} t)$, with
$H_{\rm loc}=\sum_{j,d(i,j)\leq l_{trunc}-R} H_j$.  Then, $O_i^{\rm trunc}(t)$
only involves sites within distance $l_{trunc}$ of $i$ (assuming that $O_i$
originally involved only sites within that distance).  Then,
\begin{eqnarray}
\label{truncb}
||O_i(t)-O_i^{\rm trunc}(t)||\\ \nonumber
\leq \sum_{j,l_{trunc}-R<d(i,j)\leq l_{trunc}+R}
||O|| g(t,l_{trunc})\\ \nonumber
\equiv N(l_{trunc}) ||O|| g(t,l_{trunc}),
\end{eqnarray}
where $N(l_{trunc})$ is
defined to be the number the number of sites $j$ with
$l_{trunc}-R<d(i,j)\leq l_{trunc}+R$.

In Eq.~(\ref{dmi}), for $|t|<c_1 l_{proj}$ we can use the
bound on $g$ to bound the difference $||H_i(t)-H_i^{\rm trunc}(t)||$, while
for $|t|>c_1 l_{proj}$ we use $||H_i(t)-H_i^{\rm trunc}(t)||\leq 2 J$.
Thus,
\begin{eqnarray}
||\tilde H_i^0-M_i|| \leq
J (N(l_{proj}) g(c_1 l_{proj},l_{proj}) + \\ \nonumber
2 \exp[-(c_1 l_{proj} \Delta E)^2/(2 q)]),
\end{eqnarray}
giving Eq.~(\ref{apb}).

\section{Commutator in Low Energy Sector}
Here we consider sufficient conditions for the $M_i$ to commute
in the low energy sector.  We consider the case that $n$ is
uniformly bounded above, and $\Delta E$ is uniformly bounded below,
independent of system size.
The important result is that for
most systems of interest, including all translationally
invariant systems, the commutator of the $M_i$ is exponentially
small in the system size; for a few examples certain of the $M_i$
do not commute, but even in that case the lack of commutation only
poses problems in a finite region of the system (in this case, as we will
see, there are local zero energy degrees of freedom).

The operators $M_i$ are local, in that each acts only within
a finite range $l_{proj}$ of site $i$.  Consider then the
operators $\tilde M_i^0$, for some $q$.  This $q$ is not necessarily
the $q$ of Eq.~(\ref{tnd}).  Using the locality results, we have
\be
\label{lc}
||[\tilde M_i^0,M_j]|| \leq
J^2 (N_M g(c_1 l,l) +
2 \exp[-(c_1 l \Delta E)^2/(2 q)]),
\ee
where $l=d(i,j)-2 l_{proj}$ is
the smallest distance between sites acted on by $M_i$ and $M_j$ and
$N_M$ is the total number of sites acted on by $M_j$.

Now, $P_{low} \tilde M_i^0 P_{low}=P_{low} M_i P_{low}$.  Therefore,
$[P_{low}\tilde M_i^0 P_{low},P_{low} M_j P_{low}]=
[P_{low} M_i P_{low},P_{low} M_j P_{low}]$.  The second commutator
is the low energy commutator that we wish to evaluate.  
We can bound the first commutator by a triangle inequality,
\begin{eqnarray}
\nonumber
&&||[P_{low}\tilde M_i^0 P_{low},P_{low} M_j P_low]|| 
\leq
||[\tilde M_i^0,M_j]||\\ \nonumber
&+& 
(||P_{high} \tilde M_i^0 P_{low}||
+||P_{low} \tilde M_i^0 P_{high}||)
||M_j||.
\end{eqnarray}
Following
Eq.~(\ref{gs}),
$||P_{high} \tilde M_i^0 P_{low}||\leq J \exp(-q/2)$ and also
$||P_{low} \tilde M_i^0 P_{high}||\leq J \exp(-q/2)$.
Combining this with Eq.~(\ref{lc}),
\begin{eqnarray}
\nonumber
||[P_{low} M_i P_{low},P_{low} M_j P_low]||
\leq
J^2 (N_M g(c_1 l,l) \\ \nonumber
+
2 \exp[-(c_1 l \Delta E)^2/(2 q)]+2 exp[-q/2]).
\end{eqnarray}
Pick $q=c_1 l \Delta E$, we find that
the commutator of
\begin{eqnarray}
\label{blecm}
||[P_{low} M_i P_{low},P_{low} M_j P_low]||\\ \nonumber
\leq
J^2 (N_M g(c_1 l,l) +
4 \exp[-c_1 l \Delta E/2]).
\end{eqnarray}
Thus, the
commutator of $M_i$ with $M_j$ in the low energy sector is
exponentially small in $d(i,j)$.

For a translationally invariant, $d$-dimensional,
system on a lattice, this suffices to show that the commutator
of $M_i$ with $M_j$ in the low energy sector is exponentially
small in the system size.  
We can introduce coordinates for each site:
 $(\vec x,y)$.
The $d$-dimensional vector $\vec x$ labels the particular unit cell
to which the site belongs, while the coordinate $y$
indicates the particular site within
that cell.  
The Hamiltonian is invariant under translation so we can pick a basis in
which the $n$ ground states are eigenvectors of the translation operators.
In this basis, the matrix element elements of the $M$ are
related by $(M_{\vec x,y})_{ab}=(M_{\vec x+\vec x',y})_{ab}
\exp[i \vec x' \cdot (\vec k_a-\vec k_b)]$, where $\vec k_a$ is the
momentum of ground state $a$, for $0\leq a \leq n-1$.

In the simplest case (a case which applies to every system of which we are
aware), the $\vec k_a$ are vectors of rational multiples $p/q$ of $2\pi$,
such that $p,q$ are integers with some denominator $q$ which is independent
of system size for large enough systems.  That is, 
$\vec k_a=2\pi (p^1_a/q^1_a,...,p^d_a/q^d_a)$.
Then, it is possible to identify a supercell, such that all
of the $n$ ground states are unchanged by translation by a supercell.
The size of this supercell in a given direction $m$ is equal to
the least common multiple of the $n$ different $q^m_a$.
For example, in the Majumdar-Ghosh chain, the two lowest states have
momentum $0$ and $\pi$, so that in one case $p=0,q=1$, and in the other
$p=1,q=2$.  Then, the ground states are unchanged under translation by
two sites.  For a system with momenta $0,\pi/3,2\pi/3$ for the ground
states, the ground states would be unchanged under translation by three
sites.  Then, consider any commutator $[M_{\vec x_1,y_1},M_{\vec x_2,y_2}]$.
This equal $[M_{\vec x_1,y_1},M_{\vec x_2+\vec x',y_2}]$ where $\vec x'$
translates $\vec x_2$ by some number of supercells.  By choosing
$\vec x'$ correctly, we can make the distance between $(\vec x_1,y_2)$ and
$(\vec x_2+\vec x',y_2)$ of order the linear size of the system.
Then, we can use the result that the commutator vanishes exponentially
in the spacing between the operators to show that
the original commutator $[M_{\vec x_1,y_1},M_{\vec x_2,y_2}]$ is exponentially
small in the system size, as desired\cite{self}.

A slightly more complicated, but very artificial, case, is that in which
the $p,q$ depend on system size in such a way that the size of the supercell
is equal to the linear size of the system.  It is not clear that such
a thing can actually happen in a system with a finite number $n$ of
low energy states.  Even in this case it is possible to show
the commutator of any two $M_i,M_j$ in the low energy sector
is exponentially small in the system size, by using the
smallness of the commutator $[M_i,M_j]$ for large $d(i,j)$ and
expanding the commutator in intermediate states.  However, the proof is
sufficiently artificial that we do not give it here.

What if a system does not have translational invariance?  In this case,
the $M_i$ need not commute in
the low energy sector.  Consider the following example
system, a one-dimensional system of spin-$1/2$ on each site $i$.
For $i\neq 0,1,2$, we have $H_i=S^z_i$, while $H_0=S^x_0+S^y_0$,
$H_1=-S^x_0-S^z_1$, and $H_2=-S^y_0+S^z_2$.  This is a complicated
way of writing the Hamiltonian $H=\sum_{i\neq 0} S^z_i$, since all the
terms acting on site $0$ cancel between $H_0$, $H_1$, and $H_2$.
This Hamiltonian has a doubly degenerate ground state, with all spins
pointing down, except for spin $0$ which can point in either direction.
For large enough $q,l_{proj}$, the commutator of the $M_i$ in the
low energy sector is close to that of the $H_i$ in the low energy
sector.
However, clearly $H_0$ and $H_1$ do not commute in the low energy
sector, since $[S^y_0,S^x_0]\neq 0$.  Similarly, $H_1$ and $H_2$ do not
commute, and $H_0$ and $H_2$ do not commute.  Still $H_0$ commutes in
the low energy sector with $H_i$ for $i\geq 2$.

In this case, the lack of commutation is localized near site $0$.
The two ground states differ only locally, on site $0$, and only
the $H_i$ for $i=0,1,2$ fail to commute, while the others are
diagonal in the low energy subspace.

We now show how to choose
a basis of the ground states
to bound $\sum_i o(i)$.
Consider a system with $n=2$.  Then, find
the $j$ which maximizes $\delta\lambda_j$, where
$\delta\lambda_j$ is the difference between the two
eigenvalues of $M_j$
in the low energy
sector, and work in a basis which diagonalizes
$M_j$ in the low energy sector.  Then, $o(i)\leq
||[P_{low} M_i P_{low},P_{low} M_j P_{low}]||/(\delta\lambda_j)$.  From
Eq.~(\ref{blecm}), the commutator is an exponentially decaying function
of $d(i,j)$, and summing over all $i$ we find that
\be
\label{oibu}
\sum_i o(i) \leq {\cal O}(J^2/c_1 l\Delta E\delta\lambda_j),
\ee
independent
of system size.

For a system
with $n>2$, we can proceed similarly; we first find the $M_j$ with the
maximum difference between its largest and smallest eigenvalues, and
and use this to bound the off-diagonal
matrix elements of other $M_i$ between the corresponding eigenvectors.
We then find an $M_k$ which has two different eigenvalues, with different
eigenvectors from $M_j$, and show that off-diagonal
matrix elements between those two states are exponentially small in
$d(i,k)$.  
We proceed like
this until we have bounded all off-diagonal matrix elements.


\begin{thebibliography}{99}
\bibitem{split} The case of almost degenerate low energy states, with
a splitting which is very small, say exponentially small in the system
size, can be handled with only very minor changes to the calculations
in this paper.
\bibitem{exp} One can also consider interactions which decay exponentially
in space.
\bibitem{notation} We use the notation $\langle \Psi_1 , \Psi_2 \rangle$ for
the inner product, rather than $\langle \Psi_1 | \Psi_2 \rangle$, to avoid
confusion with the use of $|...|$ to denote a norm.
\end{thebibliography}
\end{document}